\documentclass[aps,prl,showpacs,twocolumn]{revtex4}
\usepackage{graphicx}
\input{epsf}

\begin{document}
\title {
Quantum dissociation
of a vortex-antivortex pair in a long Josephson
junction}

\author{M. V. Fistul$^1$, A. Wallraff$^{1,2}$, Y. Koval$^1$,
A. Lukashenko$^1$, B. A. Malomed$^3$,
and A. V. Ustinov$^1$}

\affiliation {$^1$Physikalisches Institut III,
Universit\"at Erlangen-N\"urnberg, D-91058, Erlangen, Germany}
\affiliation {$^2$ Department of Applied Physics,
Yale University, New Haven, CT 06520, USA}
\affiliation {$^3$ Department of Interdisciplinary Science, Faculty of Engineering,
Tel Aviv University, Tel-Aviv 69978, Israel}
\date{\today}
\begin{abstract}
We report a theoretical analysis and experimental observation of the quantum dynamics of a single
vortex-antivortex (VAV) pair confined in a long narrow annular Josephson junction.
The switching of the junction from the superconducting state to the 
resistive state occurs
via the dissociation of a pinned VAV pair.
The pinning potential is controlled by external magnetic field $H$
and dc bias current $I$.
We predict a specific magnetic field dependence of the oscillatory energy levels of the
pinned VAV state and the crossover to a {\it macroscopic quantum
tunneling} mechanism of VAV dissociation at low temperatures.
Our analysis explains the experimentally observed {\it increase}
of the width of the switching current distribution $P(I)$ with $H$ and the
crossover to the quantum regime at the temperature of about $100$ mK.
\end{abstract}


\maketitle

Great attention has been devoted to the experimental and
theoretical study of {\it macroscopic quantum phenomena} in
diverse Josephson coupled systems \cite{Tinkham,Clarke1,WalUst,HanMartMoya}.
Most of these systems, e.g. dc biased single Josephson junctions (JJs), various SQUIDs and
small Josephson junction arrays, contains a few {\it lumped} Josephson junctions, hence, they
can be described by a few degrees of freedom (Josephson phases).
At low temperature quantum-mechanical effects such as macroscopic tunneling, energy levels
quantization \cite{Clarke1,WalUst}, and coherent oscillations \cite{HanMartMoya} of the
Josephson phase have been observed.

As we turn to {\it spatially extended} Josephson systems, including quasi-one-dimensional
long Josephson junctions, parallel arrays and Josephson junction ladders, which 
present a particular case of interacting many particle systems, the analysis and observation
of macroscopic quantum dynamics is more complex.
These systems can support diverse nonlinear excitations,
such as Josephson vortices (magnetic fluxons) and vortex-antivortex (VAV) pairs, which interact with
inhomogeneities and linear (Josephson plasma) modes \cite{UstRev}.
The classical dynamics of such excitations is well established, and in particular,
the thermal fluctuation induced escape of a Josephson phase 
from the metastable state has been studied \cite{Chirillo,EscapeCl}. 
It is necessary to stress that, while various macroscopic quantum-mechanical effects 
have been predicted  
\cite{KIM,ShnBenMal,WalLowTemp,Maki,MKR,Kato}, only few of them
were observed in experiments \cite{Moya1,EscapeQn}.
In particular, studying the Josephson phase escape from the metastable state, 
the macroscopic quantum tunneling of a state with many vortices has been observed in
\cite{Moya1}, and recently tunneling of a single vortex  and its energy level quantization
have been measured \cite{EscapeQn}.

\begin{figure}
\includegraphics[width=2.3in]{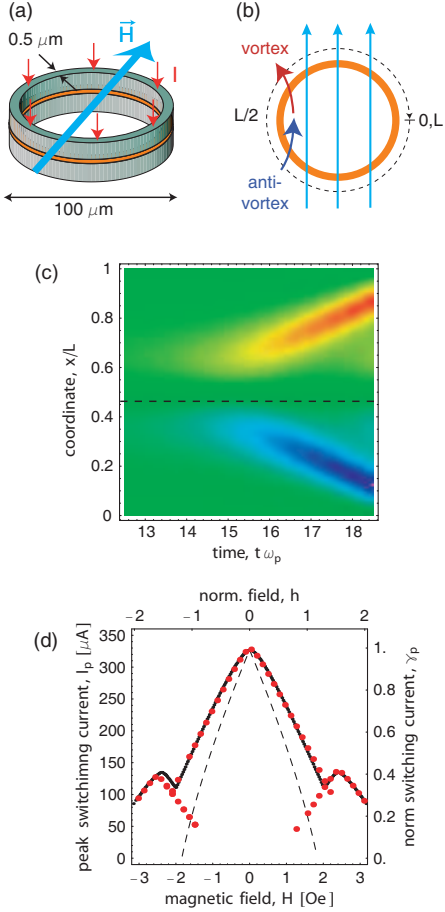}
\caption{(a) A schematic view of the system. A long
annular Josephson junction \emph{without any trapped vortices} is
placed in the in-plane external magnetic field $H$ and biased by 
uniform current $I$. (b) Generation of a confined vortex-antivortex pair with the center at $x=L/2$.
(c) Numerically simulated
evolution of the magnetic field distribution, as the Josephson junction switches
into the resistive state. The emerging vortex and antivortex moving in
opposite directions are seen. (d)
The magnetic field dependence of the
critical current for the annular junction with $L\equiv \pi d /\lambda_J =10.5$: 
experimental data measured at $T = 850$ mK (circles), numerically calculated $I_c(H)$
(dotted line), and $I_c(H)$ predicted by Eq. (\ref{critcurr-Gen})
(dashed line).
}
\label{schematic}
\end{figure}

After the quantum dynamics of a single vortex, the next step in the study of macroscopic quantum 
effects is the dynamics of a {\it single vortex-antivortex pair}. States containing many VAV pairs 
are relevant to thin superconducting films or large two-dimensional Josephson arrays
close to the Kosterlitz-Thouless transition \cite{Tinkham}. A
{\em single} VAV pair naturally appears in a long  \emph{annular} 
JJ placed in the external magnetic field $H$ parallel to the junction plane
\cite{GrEn,MM} (Fig.~\ref{schematic}a).

In this Letter, we report a theoretical analysis and experimental observation of
quantum-mechanical effects in the dynamics of a single vortex-antivortex pair.
We show that in the presence of an externally applied magnetic field $H$ a
single VAV pair can be nucleated in a long JJ. A pair is confined in a potential well
created by externally applied magnetic field and dc bias current.
Fluctuations, thermal and quantum, induce the
internal oscillations of the pair. The switching of the JJ from the
superconducting state to the resistive state occurs by {\it dissociation}
of a pinned pair. At high
temperature the dissociation takes place in a form of
thermal activation over the barrier. At low temperature there occurs {\it
macroscopic quantum tunneling} through the barrier.

First, we quantitatively analyze the penetration and following dissociation of a
VAV pair in the presence of a small magnetic field $H$ and a large dc bias, namely
$\delta~=~\frac{I_{c0}-I}{I_{c0}}~\ll~1 $, where
$I_{c0}$ is the critical current of a long JJ in
for $H=0$. In this case the Josephson phase is written as
$\varphi(x,t)=\frac{\pi}{2}+\xi(x,t)$, where a small variable part
($\xi(x,t)~\ll~1$) satisfies the equation:
\begin{equation} \label{GenEq}
 \xi_{tt}-\xi_{xx}
-\frac{\xi^2}{2}~=~-\delta-h\cos\left(\frac{2\pi x}{L}\right)~,
\end{equation}
where $L$ is the junction length, and $h~\propto~H$ is the normalized
external magnetic field. Here, 
coordinate $x$ and time $t$  are normalized to the Josephson
penetration length $\lambda_J$ and the inverse plasma frequency
$\omega_p^{-1}$, respectively.
A particular solution of this equation satisfying the boundary conditions
$\xi_x=0$ at $x=\pm \infty$ is written in the form
\begin{equation} \label{PatSol}
\xi(|x-x_1|,A)=\sqrt{2\delta} \left[\frac{3}{\cosh^2
\left(\frac{|x-x_1|+A}{2} (2\delta)^{1/4}\right)}-1\right]~.
\end{equation}
Here, $x_1(t)$ is the center of a confined pair, and
the parameter $A(t)$ which will be allowed to vary in time, determines
the distance between vortex and antivortex.

The magnetic field and dc bias create a pinning potential for such a state.
Assuming that the JJ length $L$ is much larger than the size of the pair
(which is $\simeq~\delta^{-1/4}$), we substitute (\ref{PatSol}) in the Hamiltonian 
of the underlying sin-Gordon model and minimize it with respect to $x_1$, which readily amounts 
to setting $x_1=L/2$. 
Then, we find an effective energy of the JJ as a function of $A$ 
($\dot A $ stands for the time derivative),
$$
E(A)= \frac{m_{eff}(A)}{2}{\dot A}^2 +U_{pot}(A)~~,
$$
\begin{equation} \label{energy1}
U_{pot}(A)~=~m_{eff}(A)-12 (2\delta)^{1/4}h\tanh \left(\frac{A
(2\delta)^{1/4}}{2}\right) ~~,
\end{equation}
where the effective mass of the VAV pair is 
\begin{equation} \label{effmass}
m_{eff}(A)~=~18(2\delta)^{3/2}\int_{-\infty}^A dx \frac{\sinh^2
\left(\frac{x(2\delta)^{1/4}}{2}\right) }{\cosh^6
\left(\frac{x(2\delta)^{1/4}}{2}\right) }~.
\end{equation}

The critical current in the presence of an external magnetic field
is found by minimization of the energy $E(A)$ in $A$.
Then, the critical current is found as the bias current at which the static solution
disappears,  
which assumes that (in the absence of fluctuations) the confined pair
dissociates into a set of free vortex and an antivortex moving in opposite
directions. 
After some algebra, the critical current is found to be $\delta _{c}(h)~=2h/3
$, and the corresponding critical value $A_{0}$ of $A$ being determined by the
condition $\sinh \left( (2\delta )^{1/4}A_{0}/2\right) =1$. Note that the
critical current decreases linearly with the magnetic field, in contrast to
the case of \emph{linear} long JJs, where similar consideration yields 
$\delta _{c}^{\mathrm{lin}}(h)~\propto ~h^{4/3}$. Both results are valid for
the ideal uniform bias-current distribution and $L/2\pi \gg 1$. 

Next, we performed direct numerical simulations for the finite length JJ. 
The numerically found magnetic field dependence of the critical current is shown in Fig. 1d
(dotted line). The simulations clearly show the nucleation and 
subsequent dissociation of the VAV pair (see, Fig. 1c). However, 
there is a discrepancy between the analysis and numerics in the values of $I_c$ (see, Fig. 1d).
Analysis and numerics are in a good accord as the length of a Josephson junction is increased,
$L~\geq~20$.
The lower branch of the critical current (see Fig. 1d (circles)) corresponds
to the penetration of a well separated vortex and antivortex in the junction
($A~\simeq~L/2$).
The quantum dynamics of such a state will be discussed elsewhere.

In the presence of thermal or quantum fluctuations the
dissociation of the pinned VAV pair occurs at a random value of dc bias,
i.e. at $\delta~\geq~\delta_c(h)$. Assuming 
weak fluctuations, $\delta-\delta_c(h)~\ll~\delta_c(h)$, we expand
the energy of the state around $A=A_0$ ($\delta A~=~A-A_0 $) as
\begin{equation} \label{energy2}
E(A)~=~ \frac{\chi h^{5/4}(\delta \dot A ) ^2}{2}
+\frac{3^{3/2}\sqrt h}{2}\left(\delta-\delta_c(h)\right)(\delta A)-
\frac{h^2}{6}(\delta A)^3~~,
\end{equation}
where $\chi$ is the numerical coefficient of order one. Thus, the
problem of fluctuation induced dissociation of a confined VAV pair is
mapped to a well known problem of particle´s escape from a
cubic potential. The probability of the
dissociation depends on the height of the effective potential barrier
\begin{equation} \label{PotEFF}
U_{eff}(\delta)~=~2~\cdot 3^{5/4} h^{-1/4}(\delta-\delta_c(h))^{3/2}~~.
\end{equation}
At high temperatures, the dissociation is driven by
thermal activation over this barrier. Using the known theory
describing the particle escape from such a potential well \cite{Tinkham,Clarke1}, we
find the switching rate  of a long Josephson junction
from the superconducting state to the resistive state, $\Gamma_T(I)$,
\begin{equation} \label{Tempfluct}
\Gamma_T(I)~\propto~\exp{\left[-\frac{ 2 ~\cdot 3^{5/4}h^{-1/4}
\left(\delta-\delta_c(h)\right)^{3/2} }{k_B T}\right]}.
\end{equation}
Thus, at high temperature the standard deviation of the critical
current $\sigma$ should increase with temperature and weakly depends on
the magnetic field: $\sigma_T~\propto~T^{2/3} h^{1/6}$.
Notice that $\sigma_T$ {\it increases} with $H$, in contrast to the
behavior of a small Josephson junction where
$\sigma_T~\propto~\left(I_c(H)\right)^{1/3}$ decreases with $H$.

At low temperatures the
dissociation of the VAV pair occurs through a {\it macroscopic quantum
tunneling} process. Such a quantum regime is realized at
$T<T_{cr}$, where the crossover temperature $T_{cr}$ is determined
by the frequency $\omega(\delta)$ of small oscillations
of VAV pair,
$T_{cr}~\simeq~\frac{\hbar\omega(\delta)}{2\pi k_B}~.$
In the quantum regime the frequency
$\omega(\delta)~=~\frac{3^{3/8}}{\sqrt{\chi}}
\left(\delta-\delta_c(h)\right)^{1/4}$ determines the oscillatory energy
levels $E_n$ ($n~=~0,1,2...$) of the pinned VAV state, 
$E_n~\simeq~\hbar \omega(\delta) (n+1/2)$.

Neglecting the dissipative effects, in the quantum regime the
switching rate $\Gamma_Q(I)$ of the under-barrier dissociation can be estimated, as usual, in the 
WKB approximation, which yields
\begin{equation} \label{Qfluct}
\Gamma_Q(I)\propto e^{-\frac{36U_{eff}(\delta)}{5\hbar \omega(\delta)}}
 ~=~\exp\left[ {\frac{36\sqrt{\chi}h^{-1/4}
\left(\delta-\delta_c(h)\right)^{5/4}}{5~\cdot 3^{3/8} \hbar} }\right]~.
\end{equation}
In this limit the standard deviation of the critical current
$\sigma$ is independent of temperature and, (similar to the high
temperature case), it weakly increases with magnetic field:
$\sigma_Q~\propto~ h^{1/5}~.$

Notice, that the process described above of the penetration and
subsequent dissociation of the VAV pair is based on an assumption
that the pair size is small with respect to the JJ
length. We find that the Josephson phase escape in the form of
the dissociation of the pair occurs  (in normalized units) as
$h~\geq~ \frac{3}{4} (L/2)^{-4}~$.
In the opposite limit
of very small magnetic field $h\lesssim (L/2)^{-4}$ the Josephson
phase escape occurs homogeneously in a whole junction \cite{comment}.

\begin{figure}
\includegraphics[width=2.3in]{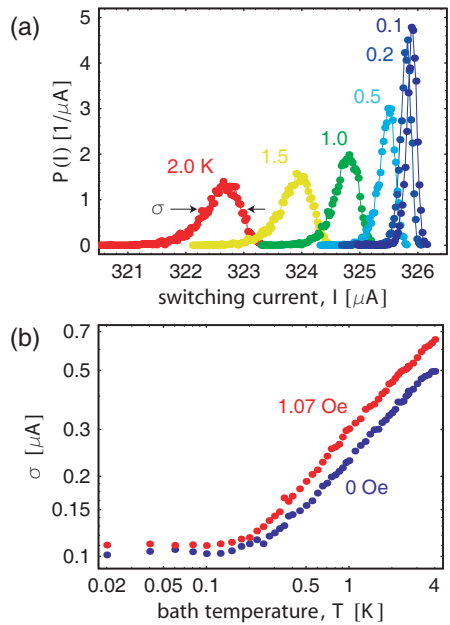}
\caption{(a) Switching current distributions $P(I-\langle I \rangle)$
at $I_H=35\,\mu$A shown at different bath temperatures $T$. The
data are plotted relative to the mean value $\langle I \rangle$ of
the switching current at each temperature. (b) Standard deviation $\sigma$ of $P(I)$ distributions
versus bath temperature. Different colors correspond to different
values of magnetic field.} \label{PI-T}
\end{figure}

The analysis presented above is valid for small
magnetic fields, $h~\ll~1$. As the magnetic field $h$ increases,
the critical current $I_c(h)$ is suppressed, and only a
{\it qualitative} description of the VAV pair dissociation can be
carried out. In the general case the pair size is
$l_b~\simeq~(1-(I/I_{c0})^2)^{-1/4}$ (instead of
$l_b~\simeq~\delta^{-1/4}$ that is valid
for $h~\ll~1$), and the amplitude of the
state is $\xi_b~\simeq~ \arccos (I/I_{c0}) $ (instead of
$\xi_b~\simeq~\sqrt{\delta}$).
Following a similar procedure as above (see, Eqs. (\ref{energy1}) and (\ref{effmass}) ) we obtain
the standard deviation dominated by thermal fluctuations
\begin{equation} \label{sigmatemp-Gen}
\sigma_T~\simeq~\frac{T^{2/3}h^{2/3}}{ \arccos \left(\frac{I_c(h)}{I_{c0}} \right)}~~,
\end{equation}
and in the quantum regime
\begin{equation} \label{sigmaquant-Gen}
\sigma_Q~\simeq~\frac{h}{\left(\arccos \frac{I_c(h)}{I_{c0}} \right)^{8/5}}~~.
\end{equation}
The standard deviation is determined by the magnetic field dependence of
the critical current $I_c(h)$ that is given implicitly by equation (this dependence is shown in
Fig. 1d by a dashed line):
\begin{equation} \label{critcurr-Gen}
h~=~\frac{3}{4}\sqrt{(1-(I_c(h)/I_{c0})^2)}\arccos \left(\frac{I_c(h)}{I_{c0}} \right)~~.
\end{equation}
A most important consequence of this analysis is the {\it saturation} of
$\sigma_{T(Q)}$ at moderate magnetic fields. The calculated dependencies $\sigma_{T(Q)}(h)$ are shown
in Fig. 3 by dashed lines.

Next, we turn to an experimental study of Josephson phase escape in
long annular JJs subject to an in-plane magnetic
field. The junction of diameter $d=100\,\mu$m and width
$w=0.5\,\mu$m was etched from a sputtered Nb/AlO$_x$/Nb thin film
trilayer and patterned using electron-beam lithography
\cite{Technology}. Having a relatively small junction width $w$ is essential for the
observation of quantum effects, as the effective mass given by Eq. (\ref{effmass}) (in natural units)
is proportional to $w$.
The critical current density of the junction
was about $220~$A/cm$^2$, which corresponds to $\lambda_J\approx
30\,\mu$m and $L\equiv\pi d/\lambda_J\approx 10.5$. The measurement
of the critical current versus magnetic field shows that no
Josephson vortices are trapped in the junction, see
Fig.~1d. The magnetic field dependence of the
critical current displays a linear decrease in a broad range of
magnetic fields, which is consistent with the analysis presented
above.

\begin{figure}
\includegraphics[width=2.3in]{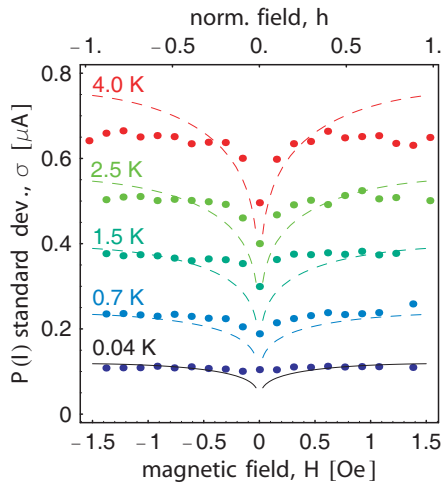}
\caption{ Standard deviation $\sigma$ of $P(I)$ distribution versus
magnetic field $H$ in the temperature range between $40$ mK and
$4.0$ K: experiment (dots), theory (dashed line)s. 
}\label{fig-sigma-H}
\end{figure}

We have measured the temperature and magnetic field dependence of
the switching current distribution $P(I)$, see Fig.~\ref{PI-T}. 
The details of measurements have been discussed elsewhere \cite{Chirillo,WalLowTemp,EscapeCl,EscapeQn}. 
As shown in Fig.~\ref{PI-T}, the width of distribution
$\sigma$ decreases with temperature and saturates below $100$
mK. This behavior indicates that at low temperatures 
the VAV dissociation in the quantum regime is observed.

For each temperature, the  distribution width
$\sigma$ has a minimum at zero magnetic field. Its magnetic field dependence
displays a peculiar \emph{increase} and a following
{\it saturation}  with magnetic
field, see Fig.~\ref{fig-sigma-H}, both in thermal and quantum
regimes. This behavior is characteristic for fluctuation
induced dissociation of the VAV state. The $\sigma(h)$
dependence is in a qualitative agreement with the theoretical
analysis given by Eqs.~(\ref{sigmatemp-Gen}) and (\ref{sigmaquant-Gen}).

In conclusion, we have shown that in the presence of a magnetic field
the switching of a long Josephson junction from the superconducting state to
the resistive one occurs through the
nucleation and subsequent dissociation of a single vortex-antivortex pair.
At low temperatures we observe the pair dissociation,
in the form of the tunneling under the barrier, in the direct experiment. 
The oscillatory energy levels of the pair
can be further studied experimentally by applying an external
resonant microwave radiation, similar to the case of a single Josephson
vortex trapped in long Josephson junction \cite{EscapeQn,FistUst}.

We would like to thank A. Kemp for technical help and A.
Abdumalikov for useful discussions. We acknowledge the partial
financial support of this project by the Deutsche
Forschungsgemeinschaft (DFG).


\begin{references}
\bibitem{comment} We do not consider here a very long Josephson junctions, 
where even in the absence of magnetic field fluctuation
induced vortex-antivortex pair are generated in the Josephson junction \cite{Maki,MKR,Kato}.
\end{references}
\end{document}